\documentclass[prd,twocolumn,eqsecnum,showpacs,amsfonts,amssymb]{revtex4}

\usepackage{graphicx}

\usepackage{bm}

\newcommand{\beq}{\begin{equation}}
\newcommand{\eeq}{\end{equation}}
\newcommand{\beqs}{\begin{eqnarray}}
\newcommand{\eeqs}{\end{eqnarray}}

\begin{document}

\title{New Scheme Transformations and Application to Study Scheme Dependence of
an Infrared Zero of the Beta Function in Gauge Theories}

\author{Gongjun Choi and Robert Shrock}

\affiliation{C. N. Yang Institute for Theoretical Physics \\
Stony Brook University \\
Stony Brook, NY 11794 }

\begin{abstract}

We present two new one-parameter families of scheme transformations and apply
these to study the scheme dependence of the infrared zero in the beta function
of an asymptotically free non-Abelian gauge theory up to four-loop order. Our
results provide a further quantitative measure of this scheme dependence,
showing that for moderate values of the gauge coupling and the parameter
specifying the scheme transformation, this dependence is relatively mild. We
also remark on a generalized multi-parameter family of rational scheme
transformations.

\end{abstract}


\maketitle


\section{Introduction}
\label{intro}

The dependence of the interaction coupling of a quantum field theory on the
Euclidean momentum scale, $\mu$, where it is measured, is of fundamental
importance.  This dependence is described by the renormalization-group beta
function of the theory \cite{rg}.  In particular, it is of interest to study
the evolution of the running gauge coupling $g \equiv g(\mu)$ of an
asymptotically free gauge theory from the deep ultraviolet (UV) region at large
$\mu$, where it is small, to the infrared (IR) region at small $\mu$.  Let us
consider such a theory (in $d=4$ spacetime dimensions) with a non-Abelian gauge
group $G$ and $N_f$ massless fermions in a given representation $R$ of $G$. If
the beta function of this theory has a zero at a value $\alpha_{IR}$, where
$\alpha=g^2/(4\pi)$, then, as the scale $\mu$ decreases from large values, the
coupling evolves toward $\alpha_{IR}$, which is thus an exact or approximate
infrared fixed point (IRFP) of the renormalization group.

The perturbative calculation of the value of $\alpha_{IR}$ at $\ell$-loop order
is complicated by the property that the terms in the beta function with $\ell
\ge 3$ depend on the scheme used for the regularization and renormalization of
the theory.  The presence of scheme dependence in higher-loop calculations is,
of course, a general property of quantum field theory; here we focus on its
effects on $\alpha_{IR}$. It is important to determine how sensitively
$\alpha_{IR}$ depends on the scheme used for its calculation. To do this, one
can compute the beta function in one scheme, then carry out a transformation to
a different scheme, and compare the respective values of the IR zero of the
beta functions in these schemes. A useful general framework is provided by
dimensional regularization of the Feynman integrals involved in loop
calculations \cite{dimreg}. An early scheme used dimensional regularization
combined with minimal subtraction of the poles at $d=4$ in the Euler $\Gamma$
functions resulting from the Feynman integrals \cite{ms}, and this was extended
to the widely used modified minimal subtraction ($\overline{\rm MS}$) scheme
with the subtraction of certain associated constants in the Taylor-Laurent
expansion of these $\Gamma$ functions \cite{msbar}. There has long been
interest in studying various scheme transformations to reduce higher-order
corrections in perturbative calculations in quantum chromodyamics (QCD) (e.g.,
\cite{mom}-\cite{brodskyreview}). In QCD, one studies the effect of applying
these scheme transformations in the vicinity of the UV zero of the beta
function at $\alpha=0$, the UV fixed point (UVFP) of QCD. 

 The situation is significantly different when one studies an IR zero of the
beta function away from the origin, $\alpha=0$.  Refs. \cite{scc,sch} pointed
out that there is much less freedom in constructing and applying acceptable
scheme transformations at an IR zero than there is at the UVFP at $\alpha=0$
and gave examples of several scheme transformations that are perfectly
acceptable in the vicinity of the UVFP at $\alpha=0$ in an asymptotically free
theory but exhibit unphysical, pathological properties, when applied at a
generic IRFP away from the origin.  Further studies of scheme transformations
and their application to an IRFP and IR properties of an asymptotically free
gauge theory have been carried out in \cite{sch2}-\cite{tr_gen}. In addition to
a variety of transformations to different schemes starting from the
$\overline{\rm MS}$ scheme \cite{msbar} studied in \cite{scc,sch,sch2,sch3},
these have included comparisons of results for the IRFP and IR properties in
the $\overline{\rm MS}$ scheme with results obtained with the modified
regularization-invariant, $\rm{RI}'$ scheme, and the minimal
momentum (MOM) subtraction scheme \cite{tr_riprime,tr_mom,tr_gen}.

In this paper we will construct and study two new one-parameter families of
scheme transformations, which we denote as $S_{L_r}$ and $S_{Q_r}$, where the
subscript $r$ is the respective parameter on which each transformation
depends. We show that these scheme transformations satisfy the rather
restrictive set of conditions set forth in \cite{scc,sch} to be physically
acceptable at an IR zero of the beta function at moderate coupling. Having done
this, we then apply them to study further the sensitivity of the IR zero of the
$\ell$-loop beta function of asymptotically free vectorial non-Abelian gauge
theories.  Our results provide a further quantitative measure of the scheme
dependence of the value of an IRFP and show that for moderate values of
$\alpha_{IR}$, as calculated at the $\ell$-loop level with $\ell$ up to four
loops, this dependence is relatively mild.  

This paper is organized as follows.  In Sect. \ref{methods} we discuss some
relevant background and the basic properties of scheme transformations.  In
Sect. \ref{sln} we present a new one-parameter family of scheme transformations
denoted $S_{L_r}$ and apply it to analyze the scheme dependence of the IR zero
of the beta function in an asymptotically free non-Abelian gauge theory up to
four-loop order.  In Sect. \ref{spq} we introduce a general class of
multi-parameter rational scheme transformations, denoted $S_{[p,q]}$, and in
Sect. \ref{sqr} we analyze a one-parameter family that is a member of this
class, namely $S_{[0,1]} \equiv S_{Q_r}$, and again apply this to study the
scheme dependence of an IR zero of an asymptotically free gauge
theory. Sect. \ref{comparisons} contains a comparison of some general features
of these scheme transformations with the $S_{\rm{sh}_r}$ scheme transformations
previously studied in \cite{scc}-\cite{sch3} involving a $\sinh$ transformation
function. Some remarks on other $S_{[p,q]}$ families of scheme transformations
are given in Sect. \ref{other_spq}. Section \ref{irfree} contains some remarks
on IR-free theories.  Our conclusions are given in Sect. \ref{conclusions}.
Certain auxiliary results are listed in an Appendix.


\section{Background and Methods}
\label{methods}


\subsection{Beta Function} 
\label{betasection} 

In this section we discuss some relevant background. We define $a(\mu)=a$ as 
\beq
a \equiv \frac{g^2}{16\pi^2} = \frac{\alpha}{4\pi} \ ,
\label{a}
\eeq
where here and below, the argument $\mu$ will often be suppressed in the
notation.  The beta function is $\beta_g = dg/dt$ or equivalently,
\beq
\beta_\alpha \equiv \frac{d\alpha}{dt} = \frac{g}{2\pi} \, \beta_g \ ,
\label{betadef}
\eeq
where $dt=d\ln \mu$. The function $\beta_\alpha$ has the series expansion
\beq
\beta_\alpha = -2\alpha \sum_{\ell=1}^\infty b_\ell \, a^\ell =
 -2\alpha \sum_{\ell=1}^\infty \bar b_\ell \, \alpha^\ell \ ,
\label{beta}
\eeq
where $\bar b_\ell = b_\ell/(4\pi)^\ell$.  The $n$-loop ($n\ell$) beta
function, denoted $\beta_{\alpha,n\ell}$, is obtained from Eq. (\ref{beta}) by
replacing the upper limit on the $\ell$ loop summation by $n$ instead of
$\infty$.  The $b_\ell$ for $\ell=1,2$ are independent of the scheme used for
regularization and renormalization, while $b_\ell$ with $\ell \ge 3$ are
scheme-dependent \cite{gross75}. For a non-Abelian gauge theory, the
coefficients $b_1$ and $b_2$ were calculated in \cite{b1} and \cite{b2}, while
$b_3$ and $b_4$ were calculated in the $\overline{\rm MS}$ scheme in \cite{b3}
and \cite{b4}. We denote the IR zero of the $n$-loop beta function
$\beta_{\alpha,n\ell}$ as $\alpha_{IR,n\ell} = 4\pi a_{IR,n\ell}$.


\subsection{Scheme Transformations} 
\label{scheme_trans_section}

A scheme transformation can be expressed as a mapping between $\alpha$ and
$\alpha'$ or equivalently, between $a$ and $a'$, namely 
\beq
a = a' f(a') \ . 
\label{aap}
\eeq
In the limit where $a$ and $a'$ vanish, the theory becomes free, so a scheme
transformation has no effect.  This implies the condition $f(0) = 1$. The 
functions $f(a')$ that we consider have Taylor series expansions about $a=a'=0$
of the form 
\beq
f(a') = 1 + \sum_{s=1}^{s_{max}} k_s (a')^s \ , 
\label{fapseries}
\eeq
where the $k_s$ are constants, and $s_{max}$ may be
finite or infinite.  Given the form (\ref{fapseries}), it follows that the
Jacobian
\beq
J = \frac{da}{da'}= \frac{d\alpha}{d\alpha'} 
\label{j}
\eeq
has the series expansion 
\beq
J = 1 + \sum_{s=1}^{s_{max}} (s+1)k_s(a')^s
\label{jseries}
\eeq
and thus satisfies
\beq
J=1 \quad {\rm at } \ \ a=a'=0 \ . 
\label{jacobianazero}
\eeq
The beta function in the transformed scheme is 
\beq
\beta_{\alpha'} \equiv \frac{d\alpha'}{dt} = \frac{d\alpha'}{d\alpha} \, 
\frac{d\alpha}{dt} = J^{-1} \, \beta_{\alpha} \ . 
\label{betaap}
\eeq
with the series expansion 
\beq
\beta_{\alpha'} = -2\alpha' \sum_{\ell=1}^\infty b_\ell' (a')^\ell =
-2\alpha' \sum_{\ell=1}^\infty \bar b_\ell' (\alpha')^\ell \ ,
\label{betaprime}
\eeq
where $\bar b'_\ell = b'_\ell/(4\pi)^\ell$.  Since 
Eqs. (\ref{betaap}) and (\ref{betaprime}) define the same function, one can 
solve for the $b_\ell'$ in terms of the $b_\ell$ and $k_s$.  This yields 
the results $b_1'=b_1$ and $b_2'=b_2$.  In \cite{scc,sch}, explicit expressions
were calculated for higher-loop $b_\ell'$ with $\ell \ge 3$ in terms of the 
$b_\ell$ and $k_s$.  In general, it was shown that the coefficient 
$b_\ell'$ with $\ell \ge 3$ in the transformed scheme is a linear combination 
of $b_n$ with $1 \le n \le \ell$ with coefficients that are algebraic functions
of the various $k_s$.  Some relevant results are given in the Appendix. 

Given that the $b_\ell$ for $\ell \ge 3$ are scheme-dependent, one may ask
whether it is possible to transform to a scheme in which the $b'_\ell$ are all
zero for $\ell \ge 3$, i.e., a scheme in which the two-loop $\beta$ function is
exact.  Near the UV fixed point at $\alpha=0$, this is possible, as emphasized
by 't Hooft \cite{thooft77}.  The resultant scheme, in which the beta function
truncates at two-loop order is commonly called the 't Hooft scheme
\cite{khuri}.

Ref. \cite{sch} presented an explicit scheme transformation which, starting
from an arbitrary scheme, transforms to the 't Hooft scheme. This
necessarily has $s_{max}=\infty$. However, Refs. \cite{sch,sch2} also noted
that although this scheme transformation is acceptable in the vicinity of a
zero of the beta function at $\alpha=0$ (UV zero for an asymptotically free
theory or IR zero for an infrared-free theory), it cannot, in general, be
applied to a generic zero of the beta function (IR zero of an asymptotically
free theory or UV zero of an infrared-free theory) away from
$\alpha=0$. Ref. \cite{sch3} constructed and studied a one-parameter class of
scheme transformations, denoted $S_{R,m,k_1}$ having $s_{max}= m \ge 2$, with
the property that an $S_{R,m,k_1}$ scheme transformation eliminates the
$\ell$-loop terms in the beta function of a gauge theory from loop order
$\ell=3$ to order $\ell=m+1$, inclusive and can be applied not only at a zero
of the beta function at $\alpha=0$ but also for a zero of the beta function
away from $\alpha =0$.

In order to be physically acceptable, a scheme transformation must satisfy
several conditions, as was discussed in \cite{sch}. We state these for an
asymptotically free gauge theory: (i) condition $C_1$: the scheme
transformation must map a real positive $\alpha$ to a real positive $\alpha'$;
(ii) $C_2$: the scheme transformation should not map a moderate value of
$\alpha$, for which perturbation theory may be reliable, to a value of
$\alpha'$ that is so large that perturbation theory is unreliable, or vice
versa; (iii) $C_3$: the Jacobian $J$ should not vanish (or diverge) or else the
transformation would be singular; and (iv) $C_4$: since the existence of an IR
zero of $\beta$ is a scheme-independent property of an theory, a scheme
transformation must satisfy the condition that $\beta_\alpha$ has an IR zero if
and only if $\beta_{\alpha'}$ has an IR zero. Since $J=1$ for $a=a'=0$, a
corollary of condition $C_3$ is that $J$ must be positive.  Since one can
define a transformation from $\alpha$ to $\alpha'$ and the inverse from
$\alpha'$ to $\alpha$, these conditions apply going in both directions.  In
passing, we note that with obvious changes (IR zero $\to$ possible UV zero in
condition $C_4$), these conditions also apply to to an infrared-free gauge
theory such as U(1) and a non-Abelian gauge theory with sufficiently many
fermions, as discussed in \cite{lnf}, and to an (infrared-free) scalar theory,
such as an O($N$) $\lambda|\vec{\phi}|^4$ theory, as analyzed in \cite{lam}.

These four conditions $C_1$-$C_4$ can always be satisfied by scheme
transformations used to study the UV fixed point in an asymptotically free
theory.  However, as was pointed out in \cite{scc} and shown with a number of
examples in \cite{scc}-\cite{sch3}, they are not automatically satisfied, and
indeed, are quite restrictive conditions when one applies the scheme
transformation at a zero of the beta function away from the origin, $\alpha=0$,
i.e., at an IR zero of the beta function for an asymptotically free theory or a
possible UV zero of the beta function for an infrared-free theory.  For
example, recall the scheme transformation denoted $S_{\rm{th}_r}$
\cite{scc,sch}, defined by $a = (1/r) \, \tanh(ra')$, depending on a parameter
$r$. Since this transformation is an even function of $r$, one may take $r \ge
0$ without loss of generality. The $S_{\rm{th}_r}$ transformation is
well-behaved near the UVFP at $a=a'=0$ in an asymptotically free theory, but is
not acceptable at a generic IR zero of the beta function.  The reason is
evident from its inverse, $a' = (2r)^{-1} \, \ln[(1+ra)/(1-ra)]$.  As $ra$
approaches 1 from below, $a' \to \infty$, and for $ra > 1$, $a'$ is complex.
Hence, this transformation violates conditions $C_1$, $C_2$, and $C_4$. For
example, for $r=4\pi$, this scheme transformation is $\alpha=\tanh \alpha'$ and
the inverse is $\alpha' = (1/2)\ln[(1+\alpha)/(1-\alpha)]$, with the
pathologies occurring as $\alpha$ approaches 1 from below. For $r=8\pi$, the
pathologies occur as $\alpha$ approaches the value 0.5 from below. As this
example and the others analyzed in \cite{scc}-\cite{sch3} show, the
construction and application of a physically acceptable scheme transformation
at a zero of the beta function away from the origin is considerably more
difficult than at a zero of the beta function at the origin, as in scheme
transformations used in QCD \cite{brodskyreview}.

In the following, to avoid overly complicated notation, we will use the generic
notation $\alpha'$ for the result of the application of each scheme
transformation to an initial $\alpha$, with it being understood that this
refers to the specific transformation under consideration. Where it is
necessary for clarity, we will use a subscript to identify the specific scheme
$S$ being discussed.  


\subsection{UV to IR Evolution of Non-Abelian Gauge Theories} 

Since we will apply our new scheme transformations to study the scheme
dependence of an IR zero of the beta function for a vectorial, asymptotically
free (non-Abelian) gauge theory, it is appropriate to review briefly some of
the properties of this theory.  Let us consider such a theory with gauge group
$G$ and $N_f$ massless fermions transforming according to a representation $R$
of $G$. Our assumption of massless fermions does not entail any significant
loss of generality, since if a given fermion had a mass $m$, then in the UV to
IR evolution of the theory, as the reference Euclidean momentum scale $\mu$
decreased past $m$, one would integrate out this fermion to construct the
low-energy effective field theory applicable at scales $\mu < m$, so the
further evolution into the IR would be essentially equivalent to a theory
without this massive fermion present.  With the minus sign extracted in
Eq. (\ref{beta}), the asymptotic freedom of the theory means that the one-loop
coefficient $b_1$ in Eq. (\ref{beta}) is positive. As $N_f$ increases, $b_1$
decreases and eventually would vanish at $N_{f,b1z} = 11C_A/(4T_f)$ \
\cite{casimir,nfreal}. Thus, the asymptotic freedom yields an upper bound on
$N_f$, namely, $N_f < N_{f,b1z}$. 

For small $N_f$, the two-loop coefficient $b_2$ has the same positive sign as
$b_1$, so the (perturbatively calculated) two-loop beta function,
$\beta_{\alpha,2\ell}$, has no IR zero.  The coefficient $b_2$ decreases as
$N_f$ increases and passes through zero to negative values as $N_f$ ascends
through the value
\beq
N_{f,b2z} = \frac{17 C_A^2}{2T_f(5C_A+3C_f)} \ .
\label{nfb2z}
\eeq
Since $N_{f,b2z} < N_{f,b1z}$, there is an interval of values of $N_f$, denoted
$I$, given by
\beq
I: \quad N_{f,b2z} < N_f < N_{f,b1z} \ ,
\label{nfinterval}
\eeq
in which the two-loop beta function has an IR zero.  This occurs at
$a_{IR,2\ell}=-b_1/b_2$, i.e.
\beq
\alpha_{IR,2\ell} = - \frac{4\pi b_1}{b_2} \ , 
\label{alfir_2loop}
\eeq
which is physical for $b_2 < 0$ \cite{b2,bz}. The scheme independence of $b_1$
and $b_2$ implies that $\alpha_{IR,2\ell}$ is also scheme-independent. Since an
IR zero of $\beta_{\alpha,n\ell}$ for $n \ge 3$ depends on the scheme $S$ used
for the computation, we denote it here as $\alpha_{IR,n\ell,S}$.  

Let us assume $N_f \in I$, so that $\beta_{\alpha,2\ell}$ has an IR zero,
$\alpha_{IR,2\ell}$. If $N_f$ is close to $N_{f,b1z}$, then, as noted,
$\alpha_{IR,2\ell}$ is small. In this case, one expects that the UV to IR
evolution of the theory leads to a deconfined non-Abelian Coulomb phase without
any spontaneous chiral symmetry breaking \cite{bz}. In this case, the IR zero
is an exact IRFP. As $N_f$ decreases, $\alpha_{IR,2\ell}$ increases.  If
$\alpha_{IR,2\ell}$ is sufficiently large, the UV to IR evolution generically
leads to the formation of bilinear fermion condensates in the most attractive
channel, with attendant spontaneous chiral symmetry breaking and dynamical
generation of effective masses for the fermions involved. In the ladder
approximation to the Schwinger-Dyson equation for the fermion propagator, this
occurs as $\alpha$ increases through a value $\alpha_{cr}$ given by
\cite{chipt} $\alpha_{cr} = \pi/(3C_f)$. Taking account of the intrinsic
uncertainties involved in the strongly coupled physics of fermion condensate
formation, one may infer more generally that the actual critical value of
$\alpha$ is expected to satisfy $\alpha_{cr} C_f \sim O(1)$. The fermions
involved in the condensate gain dynamical masses of order the
chiral-symmetry-breaking scale and are integrated out of the low-energy
effective field theory below this scale. Thus, the beta function changes to one
with the effective $N_f=0$, which does not have an IR zero, and hence the gauge
coupling increases, eventually exceeding the range where perturbative
calculations are applicable. In this case, the IR zero is only an approximate
IRFP of the renormalization group.  One defines a critical value, $N_{f,cr}$,
that separates the two types of UV to IR evolution; for $N_f > N_{f,cr}$, this
evolution is to a massless non-Abelian Coulomb phase, while for
$N_f < N_{f,cr}$, it involves the above-mentioned chiral symmetry breaking.

As $N_f$ decreases toward $N_{f,cr}$, the resultant IR zero occurs at
moderately strong coupling, and consequently it is necessary to go beyond the
two-loop level and calculate $\alpha_{IR,n\ell}$ at higher loop order
\cite{gkgg}.  This was done up to four-loop order for $\alpha_{IR,n\ell}$ and
for the anomalous dimension, $\gamma_m$, of the fermion bilinear for a general
gauge group and fermion representation in \cite{bvh,ps}.  For fermions in the
fundamental representation, it was found that, in the $\overline{\rm MS}$
scheme, relative to the (scheme-independent) two-loop value,
$\alpha_{IR,2\ell}$ \ .
\beq
\alpha_{IR,3\ell,\overline{\rm MS}} \ < \ 
\alpha_{IR,4\ell,\overline{\rm MS}} \ < \ \alpha_{IR,2\ell} \ . 
\label{alfir_loop_dep_msbar}
\eeq
The shifts in the value of the IR zero with ascending loop order were found to
become smaller as $N_f$ approaches $N_{f,b1z}$.  Comparisons were made with the
extensive lattice studies of this physics for various gauge groups and fermion
representations \cite{lgtrev}.  Further higher-loop results on structural
properties of $\beta$ and application to the IRFP were calculated in
\cite{bfs}-\cite{bc}. Because the coefficients $b_\ell$ for $\ell \ge 3$ are
scheme-dependent, these higher-loop calculations naturally led to the study of
scheme-dependence in \cite{scc}-\cite{tr_gen}.  In the region of $N_f$ slightly
less than $N_{f,cr}$, where the theory confines but behaves in a
quasi-scale-invariant manner over an extended interval in $\mu$, some insight
has been gained from continuum studies of the changes in the spectrum of
gauge-singlet hadrons as compared with the spectrum in a QCD-like theory
\cite{chipt,bs}.  Intensive research on this region exhibiting
quasi-scale-invariant behavior has also considerably deepened one's knowledge
of this physics \cite{lgtrev}.

Let us consider a well-behaved (family of) scheme transformation(s) 
$S_{ \{ r \} }$ 
where in this paragraph, $\{ r \}$ symbolizes a set of one or more
parameters, such that $S_{ \{ 0 \} }$ is the identity.  It follows that if one
applies the transformation $S_{ \{ r \} }$ to the $\overline{\rm MS}$ scheme,
then, for a given loop order $n$,
\beq 
\lim_{ \{ r \} \to \{ 0 \} } \alpha'_{IR,n\ell,S_{ \{ r \} } }
= \alpha_{IR,n\ell,\overline{\rm MS}} \ .
\label{alfprime_r0}
\eeq
Furthermore, since the IR zero in $\beta_{\alpha,n\ell}$ approaches zero as 
$N_f$ approaches $N_{f,b1z}$ from below, one has the formal result that, with
$N_f$ extended from a nonnegative integer variable to a nonnegative real
variable, 
\beq
\lim_{N_f \nearrow N_{f,b1z}} \alpha_{IR,n\ell,S_{ \{ r \} } } = 
\lim_{N_f \nearrow N_{f,b1z}} \alpha_{IR,n\ell,\overline{\rm MS}} = 0 \ .
\label{alf_lim_Nfb1z}
\eeq
Moreover, if the set of parameters $\{ r \}$ specifying the scheme
transformation is such that this transformation is sufficiently close to the
identity, then it preserves the relative order of the values of the IR zeros of
the $n$-loop beta function.  We recall that for fermions in the
fundamental representation, in the $\overline{\rm MS}$ scheme, the three-loop
and four-loop values of the IR zero are in the order given by 
Eq. (\ref{alfir_loop_dep_msbar}) above.


\section{The $S_{L_r}$ Scheme Transformation}
\label{sln}

Here we introduce and study a scheme transformation, denoted $S_{L_r}$, where
$L$ stands for logarithm and $r$ for the parameter on which a transformation in
this family depends. This is thus actually a one-parameter family of scheme
transformations. We show that the $S_{L_r}$ scheme transformation satisfies the
necessary conditions to be acceptable at a zero of the beta function away from
the origin, for a reasonable range of $|r|$, and we then apply it to the
calculation, at higher-loop order, of an IR zero of the beta function for an
asymptotically free non-Abelian gauge theory.  This calculation provides a
measure of the scheme dependence of the value of this IR zero.

The $S_{L_r}$ scheme transformation is defined as
\beq
S_{L_r}: \quad a=\frac{\ln(1+ra')}{r} \ , 
\label{aap_sln}
\eeq
where $r$ is a (real) parameter. Writing Eq. (\ref{aap_sln}) in the form of
Eq. (\ref{aap}), the transformation function is
\beq
S_{L_r}: \quad f(a') = \frac{\ln(1+ra')}{ra'} \ . 
\label{fap_sln}
\eeq
This transformation function satisfies
\beq
\lim_{a' \to 0} f(a')=1 \ ,
\label{aaptozero_sln}
\eeq
in accordance with the requirement that $f(0)=1$.  Note also that
\beq
\lim_{r \to 0} f(a')=1 \ ,
\label{rtozero_sln}
\eeq
where the limit may be taken through either positive or negative values of $r$.
The scheme transformation (\ref{aap_sln}) has the inverse
\beq
a' = \frac{e^{ra}-1}{r} \ . 
\label{apa_sln}
\eeq
The Jacobian $J=da/da'$ is 
\beq
J = \frac{1}{1+ra'} = e^{-ra} \ . 
\label{j_sln}
\eeq
The transformation function
$f(a')$ has the Taylor series expansion
\beq
f(a') = 1+\sum_{s=1}^\infty \, \frac{(-ra')^s}{s+1} \ , 
\label{fap_sln_series}
\eeq
so, in the notation of Eq. (\ref{fapseries}), the expansion coefficients are
\beq
k_s = \frac{(-r)^s}{s+1} \ . 
\label{ks_sln}
\eeq
Thus, for small $|r|a'$,
\beq
a = a'\Big [ \, 1- \frac{ra'}{2} + O\Big ((ra')^2\Big ) \ \Big ] \ .
\label{aaptaylor_sln}
\eeq
It follows that with the application of the $S_{L_r}$ scheme transformation,
\beqs
S_{L_r}: \quad & & a' > a \quad {\rm if} \ r > 0 \cr\cr
               & & a' < a \quad {\rm if} \ r < 0 \ . 
\label{aap_inequality_sln}
\eeqs

The requirement that the right-hand side of Eq. (\ref{aap_sln}) be real implies
that the argument of the log must be positive, which, in turn, yields the
formal lower bound on this parameter
\beq
r > -\frac{1}{a'} \ . 
\label{rlowerbound_sln}
\eeq
This is also required by the condition $C_3$, that the Jacobian must be
(finite) and positive.  If $r > 0$, this inequality is obviously satisfied,
since $a$ and $a'$ are positive.  Let us then consider negative
$r$. Substituting Eq. (\ref{apa_sln}), the inequality (\ref{rlowerbound_sln})
becomes $r > r/(1-e^{ra})$. Since we have restricted to negative $r$, this can
be rewritten as $-|r| > -|r|/(1-e^{-|r|a})$, i.e., $1 < 1/(1-e^{-|r|a})$, which
is always satisfied.  Thus, $r$ may be positive or negative, and the actual
range of $r$ is determined by the conditions $C_1$ and $C_2$, that given a
value of $\alpha=4\pi a$ for which perturbative calculations are reasonably
reliable, the same should be true of $\alpha' = 4\pi a'$. This will be
discussed further below. 

Substituting the result (\ref{ks_sln}) for $k_s$ into the general expressions
for the $b'_\ell$ from \cite{sch}, we obtain
\beq
b'_3 = b_3 - \frac{r}{2} \, b_2 -\frac{r^2}{12}\, b_1 \ ,
\label{b3prime_sln}
\eeq
\beq
b'_4 = b_4 -rb_3+\frac{r^2}{4}\, b_2+\frac{r^3}{12}\, b_1 \ ,
\label{b4prime_sln}
\eeq
\beq
b'_5 = b_5-\frac{3r}{4}\, b_4+\frac{5r^2}{6}\, b_3-\frac{r^3}{8}\, b_2
-\frac{13r^4}{180}b_1 \ ,
\label{b5prime_sln}
\eeq
\beq
b'_6 = b_6-2rb_5+\frac{5r^2}{3}\, b_4-\frac{2r^3}{3}\, b_3
+\frac{7r^4}{120}\, b_2 +\frac{11r^5}{180}\, b_1 \ ,
\label{b6prime_sln}
\eeq
and so forth for the $b'_\ell$ with $\ell \ge 7$.

We next apply this $S_{L_r}$ scheme transformation to the beta function, in the
$\overline{\rm MS}$ scheme, of an asymptotically free gauge theory.  We take
the gauge group to be $G={\rm SU}(N)$. Since the $b_\ell$ have only been
calculated up to $\ell=4$ loops in the $\overline{\rm MS}$ scheme, we will only
need the results above for $b_3'$ and $b_4'$.  For $N_f \in I$, so the two-loop
$\beta$ function has an IR zero, we then calculate the resultant IR zero in
$\beta_{\alpha'}$ at the three- and four-loop order. We have carried out these
calculations with a range of values of $N$ and $r$. For $N_f \in I$ and various
values of $r$, we list the results for $N=3$, i.e., $G={\rm SU}(3)$, in Table
\ref{alfir_3loop_sln} for the zero of the three-loop beta function and in Table
\ref{alfir_4loop_sln} for the zero of the four-loop beta function. We denote
the IR zero of the $n$-loop beta function in the transformed scheme,
$\beta_{\alpha',n\ell}$, as $\alpha'_{IR,n\ell} \equiv
\alpha'_{IR,n\ell,S_{L_r}}$, and, to save space in the tables we further
shorten this to $\alpha'_{IR,n\ell,r}$.  Here and below, for this SU(3) theory,
the lower end of the interval $I$, namely $N=N_{f,b2z}$, is at $N=8.05$
\cite{nfreal}, so, for physical, integral values of $N_f$, it is $N_f=9$.  The
lowest value we show in Table \ref{alfir_3loop_sln} and the later tables is
$N_f=10$, because for $N_f=9$, $\alpha_{IR,2\ell}$ is too large for the
perturbative methods that we use to be reliable.  Our results for $N=2, \ 4$,
and other values are similar, so the $N=3$, i.e., SU(3) results displayed in
Tables \ref{alfir_3loop_sln} and \ref{alfir_4loop_sln} will be sufficient for
our discussion here. The range of $r$ for which we list results in these tables
is $-3 \le r \le 3$.  This range evidently satisfies the conditions
$C_1$-$C_4$.  For this range, the $S_{L_r}$ scheme transformation provides a
useful quantitative measure of the scheme dependence of the IR zero in the beta
function for this theory. Of course, if one were to increase the magnitude of
$|r|$ to excessively large values, with either sign of $r$, this scheme
transformation would not be useful, because it would violate conditions $C_1$
and $C_2$.  For example, in the SU(3) theory with the illustrative value
$N_f=12$, as one increases $r$ beyond the upper end of the range that we show,
for the values $r=4, \ 5, \ 6, \ 7$, one gets the four-loop result
$\alpha_{IR,4\ell,S_{L_r}}$ equal to 0.529, \ 0.550, \ 0.578, \ 0.618.  But for
$r=8$, the transformation yields a complex, unphysical result for
$\alpha_{IR,4\ell,S_{L_r}}$.  Similarly, for this $N_f=12$ case, as one
decreases $r$ below the lowest negative value, $r=-3$, the solution for
$\alpha_{IR,4\ell,S_{L_r}}$ decreases smoothly to 0.390 at $r=-10$, but becomes
complex for $r=-11$.  The resultant restriction on the range of the parameter
$r$ is generic. Thus, as was discussed before in \cite{scc}-\cite{sch3}, in
applying scheme transformations, one must necessarily restrict the form of the
transformation so as to satisfy the conditions $C_1$-$C_4$.

We also observe the following additional general properties in our calculations
of $\alpha'_{IR,n\ell,S_{L_r}}$.  First, it follows from (\ref{alfprime_r0})
together with the fact that Eq. (\ref{aap_sln}) is a continuous transformation,
that for small $|r|$, the relative order of the values of the $n$-loop IR zeros
of $\beta_{\alpha'}$ in the transformed scheme are the same as those in the
original $\overline{\rm MS}$ scheme, as given in (\ref{alfir_loop_dep_msbar}).
This is evident from the illustrative $N=3$ results given in Tables
\ref{alfir_3loop_sln} and \ref{alfir_4loop_sln}. In accord with
(\ref{alf_lim_Nfb1z}), the shifts of the value of the IR zero as a
function of loop order are larger for smaller $N_f$ and get smaller as $N_f$
approaches $N_{f,b1z}$.

Second, for a given $N$, $N_f \in I$, loop order $n=3$ or $n=4$, and $r$ values
for which the $S_{L_r}$ transformation satisfies the conditions $C_1$-$C_4$, 
\beq
\alpha'_{IR,n\ell,S_{L_r}} \quad {\rm is \ an \ increasing \ function \ of} \ 
r. 
\label{alfir_sln_r_dependence}
\eeq
This second property, in conjunction with the general property 
(\ref{alfprime_r0}), implies that, for a given $N$, $N_f \in I$, and $r$,
\beqs
& & \alpha'_{IR,n\ell,S_{L_r}} > \alpha_{IR,n\ell,\overline{MS}} \quad
    {\rm if} \ r > 0 \ \ {\rm and} \cr\cr
& & \alpha'_{IR,n\ell,S_{L_r}} < \alpha_{IR,n\ell,\overline{MS}} \quad
    {\rm if} \ r < 0 \ . \cr\cr
& & 
\label{sln_versus_msbar}
\eeqs
This holds for arbitrary loop order $n$ and, in particular, for the loop orders
$n=3$ and $n=4$ for which we have done calculations using the known
$\overline{\rm MS}$ beta function coefficients.  The result
(\ref{sln_versus_msbar}) is evident in the illustrative $N=3$ results given in
Tables \ref{alfir_3loop_sln} and \ref{alfir_4loop_sln}. In accord with
(\ref{alf_lim_Nfb1z}), the shifts of the value of the IR zero as a function of
$|r|$ become quite small as $N_f$ approaches $N_{f,b1z}$ from below.  In this
region, these shifts in the position of the IR zero of the $\ell$-loop beta
function in the transformed scheme can be sufficiently small that the entries
may coincide to the given number of significant figures displayed in the
tables.


\section{The Rational Scheme Transformation $S_{[p,q]}$ }
\label{spq}

In \cite{scc}-\cite{sch3} a number of scheme transformations were studied for
which the transformation function $f(a')$ has the form (\ref{fapseries}) with 
finite $s_{max}$, i.e., is a (finite) polynomial in $a'$.  One way that it is
possible to generalize these is to make $f(a')$ a rational function of $a'$,
i.e., 
\beq
a=a'f(a')_{[p,q]}
\label{aapspq}
\eeq
with
\beq
S_{[p,q]}: \quad f(a')_{[p,q]} =  \frac{{\cal N}(a')}{{\cal D}(a')} \ , 
\label{fapspq}
\eeq
where the numerator and denominator functions ${\cal N}(a')$ and ${\cal D}(a')$
are polynomials of respective finite degrees $p$ and $q$ in $a'$:
\beq
{\cal N}(a') = \sum_{i=0}^{p} u_i \, (a')^i \quad {\rm with} \ u_0 = 1 
\label{spq_numerator}
\eeq
and
\beq
{\cal D}(a') = \sum_{j=0}^{q} v_j \, (a')^j \quad {\rm with} \ v_0 = 1 \ . 
\label{spq_denominator}
\eeq
The restrictions that $u_0=v_0=1$ are imposed so that $f(a')$ satisfies the
necessary condition that $f(0)=1$.  Thus, a general $S_{[p,q]}$ scheme
transformation depends on the $p+q$ parameters $u_i$, $i=1,...,p$ and $v_j$,
$j=1,...,q$. As indicated, we label this class of scheme transformations as
$S_{[p,q]}$, with the dependence on the coefficients $u_i$ and $v_j$ kept
implicit.  If $q=0$, then this gives a Taylor series expansion
(\ref{fapseries}) of $f(a')_{[p,q]}$ with $s_{max}=p$, while if $q \ge 1$, then
$s_{max} = \infty$.

We note that, as one may recall from the theory of Pad\'e approximants, for a
given series expansion (\ref{fapseries}) calculated to a given finite order
$s_h$, it is possible to construct a set of rational functions $f(a')$ of the
form (\ref{fapspq}) whose Taylor series expansion coefficients match the given
set $k_s$, $s=1,...,s_h$.  Viewed the other way, if one starts with a set of
rational functions of the form (\ref{fapspq}), one knows that certain subsets
of these can be chosen to yield the same Taylor series expansion to a given
order $s_h$.

The scheme transformation function $S_{[p,q]}$ introduces $p$ zeros and $q$
poles, so a necessary requirement is that one must choose the coefficients
$u_i$ with $i=1,...,p$ and $v_j$ with $j=1,...,q$ such that the zeros and poles
occur away from the relevant physical region in $a$.  Obviously, scheme
transformations with polynomial transformation functions $f(a')$ are special
cases of $S_{[p,q]}$ with $q=0$.  Thus, the scheme transformation $S_1$ studied
in \cite{scc,sch} and \cite{sch3} is a special case of $S_{[p,q]}$ with $[p,q]=[1,0]$; the $S_2$ and $S_3$ transformations in \cite{scc,sch,sch2} are
special cases of $S_{[p,q]}$ with $[p,q]=[2,0]$ and $[p,q]=[3,0]$,
respectively; and the $S_{R,m}$ and
$S_{R,m,k_1}$ transformations studied in \cite{sch2,sch3} are special cases of
$S_{[p,q]}$ with $[p,q]=[m,0]$.  We proceed in the next section to study the
simplest member of the class of $S_{[p,q]}$ scheme transformations
with $q \ne 0$, namely the one with $[p,q]=[0,1]$. 


\section{The $S_{Q_r}$ Scheme Transformation}
\label{sqr}

In this section we introduce and apply a scheme transformation
that we call $S_{Q_r}$, defined as $S_{[p,q]}$ with $[p,q]=[0,1]$, 
\beq
S_{Q_r} \equiv S_{[0,1]} \ {\rm with} \ v_1 = -r \ .
\label{sqrspq}
\eeq
Thus, explicitly, 
\beq
S_{Q_r}: \quad a=\frac{a'}{1-ra'} \ , 
\label{aap_sqr}
\eeq
where $r$ is a (real) parameter, whose allowed range will be determined below.
As before, we show this satisfies the necessary conditions to be acceptable at
a zero of the beta function away from the origin for a reasonable range of
$|r|$, and we then apply it to assess the scheme dependence of the IR zero in
the beta function of an asymptotically free non-Abelian gauge theory at higher
loop order.  The transformation function corresponding to (\ref{aap_sqr}) is
\beq
S_{Q_r}: \quad f(a') = \frac{1}{1-ra'} \ . 
\label{fap_sqr}
\eeq
Clearly, $f(a')=1$ for $a'=0$ and for separately for $r=0$.
The inverse of Eq. (\ref{aap_sqr}) is 
\beq
a' = \frac{a}{1+ra} \ . 
\label{apa_sqr}
\eeq
The Jacobian $J=da/da'$ is 
\beq
J = \frac{1}{(1-ra')^2} = (1+ra)^2 \ . 
\label{j_sqr}
\eeq

The transformation function has the Taylor series expansion
\beq
f(a') = 1+\sum_{s=1}^\infty \, (ra')^s \ , 
\label{fap_sqr_series}
\eeq
so, in the notation of Eq. (\ref{fapseries}), the expansion coefficients are
\beq
k_s = r^s \ . 
\label{ks_sqr}
\eeq
Thus, for small $|r|a'$,
\beq
a=a'\Big [ \, 1+ra'+O \Big ( (ra')^2 \Big ) \ \Big ] \ . 
\label{aaptaylor_sqr}
\eeq
It follows that after application of the $S_{Q_r}$ scheme transformation,
\beqs
S_{Q_r}: \quad & & a' < a \quad {\rm if} \ r > 0 \cr\cr
               & & a' > a \quad {\rm if} \ r < 0 \ . 
\label{aap_inequality_sqr}
\eeqs

The condition $C_1$ requires that the denominator of the right-hand
side of Eqs. (\ref{apa_sqr}) be finite and positive, which
implies that the (real) parameter $r$ is bounded below according to
\beq
r > -\frac{1}{a} \ . 
\label{r_range_sqr}
\eeq
Clearly, in order for conditions $C_1$ and $C_2$ to be satisfied, $r$ cannot be
too close to saturating this lower bound.  Applying these conditions to the
original transformation (\ref{aap_sqr}) yields the formal inequality $r <
1/a'$.  However, substituting (\ref{apa_sqr}), this becomes $r < a^{-1}+r$,
which is always valid, since $a > 0$. Thus, the actual upper bound on $r$ is
determined by the conditions $C_1$ and $C_2$, that, given a value of $\alpha$
for which perturbative calculations are reasonably reliable, the same should be
true of $\alpha'$.

Inserting the result (\ref{ks_sqr}) for $k_s$ into the general expressions
for the $b'_\ell$ from \cite{sch}, we obtain
\beq
b'_3 = b_3 +r b_2 \ , 
\label{b3prime_sqr}
\eeq
\beq
b'_4 = b_4 + 2r b_3 + r^2 b_2 \ , 
\label{b4prime_sqr}
\eeq
\beq
b'_5 = b_5 +3r b_4 + 3r^2 b^3 + r^4 b_2 \ , 
\label{b5prime_sqr}
\eeq
\beq
b'_6 = b_6 +4r b_5 + 6r^2 b_4 + 4r^3 b_3 + r^4 b_2 \ , 
\label{b6prime_sqr}
\eeq
and so forth for the $b'_\ell$ with $\ell \ge 7$.  An important general 
property of these beta function coefficients resulting from the application of
the $S_{Q_r}$ scheme transformation to an arbitrary initial scheme is that 
\beq
S_{Q_r}: \quad b_\ell' \quad {\rm is \ independent \ of} \ \ b_1 \ \ 
{\rm for} \ \ell \ge 3 \ . 
\label{bell_indep_of_b1}
\eeq
The reason for this can be seen as follows.  The coefficient $b_\ell'$ with
$\ell \ge 3$ resulting from the application of a scheme transformation is a
linear combination of the $b_n$ with $1 \le n \le \ell$. The structure of the
coefficients multiplying these $b_n$ with $1 \le n \le \ell$ was discussed in
\cite{scc,sch}.  In particular, the respective coefficients of $b_1$ in the
expressions for $b_\ell'$ with $\ell \ge 3$ have the property that they vanish
if $k_s = (k_1)^s$.  This property is satisfied by the present $S_{Q_r}$ scheme
transformation, as is evident from Eq. (\ref{ks_sqr}).  For example, in the
expression (\ref{b3prime}) for $b_3'$ given in the Appendix, the coefficient of
$b_1$ is $k_1^2-k_2$, and in Eq. (\ref{b4prime}) for $b_4'$, the coefficient of
$b_1$ is $-2k_1^3+4k_1k_2-2k_3$; both of these coefficients of $b_1$ in $b_3'$
and $b_4'$ vanish if $k_s=(k_1)^s$. Similar results
hold for the $b_\ell'$ with higher values of $\ell$ that were calculated in
\cite{scc}-\cite{sch2}.

We next apply this $S_{Q_r}$ scheme transformation to the beta function in the
$\overline{\rm MS}$ scheme.  We present results in Table \ref{alfir_3loop_sqr}
for the three-loop calculation and in Table \ref{alfir_4loop_sqr} for the
four-loop calculation.  The range of $r$ that we use is $-3 \le r \le 3$.  For
the lowest two values of $N_f$, namely $N_f=10$ and $N_f=11$, and the lowest
values of $r$, namely $r=-3$, although the $S_{Q,r}$ scheme transformations
yields acceptable values of the three-loop zero, $\alpha'_{IR,3\ell,S_{Q_r}}$,
it yields complex values of the four-loop zero, $\alpha'_{IR,4\ell,S_{Q_r}}$.
To avoid these, one may restrict the lower range of $r$ to, e.g., $r=-2$ for
these values of $N_f$.  The $S_{Q_r}$ transformation obeys the conditions $C_1$
and $C_2$ for positive values of $r$ somewhat beyond the upper end of the range
that we show, but eventually, if one were to use excessively large values of
$r$, it would again fail to satisfy these. We thus restrict the range of $r$ 
over which we apply this $S_{Q_r}$ scheme transformation accordingly.

We remark on some general features of the $S_{Q_r}$ scheme transformation.  As
with the $S_{L_r}$ transformation, it follows from (\ref{alfprime_r0}) together
with the fact that Eq. (\ref{aap_sqr}) is a continuous transformation, that for
small $|r|$, the relative order of the values of the $n$-loop IR zeros of
$\beta_{\alpha'}$ in the transformed scheme are the same as those in the
original $\overline{\rm MS}$ scheme, as given in (\ref{alfir_loop_dep_msbar}).
This is evident in Table \ref{alfir_3loop_sqr} and from Table
\ref{alfir_4loop_sqr}.  Second, for a given $N$, $N_f \in I$, and $r$, we find
\beqs
& & \alpha'_{IR,n\ell,S_{Q_r}} < \alpha_{IR,n\ell,\overline{MS}} \quad
    {\rm if} \ r > 0 \ \ {\rm and} \cr\cr
& & \alpha'_{IR,n\ell,S_{Q_r}} > \alpha_{IR,n\ell,\overline{MS}} \quad
    {\rm if} \ r < 0 \quad {\rm for} \ \ n = 3, \ 4 \ . \cr\cr
& & 
\label{sqr_versus_msbar}
\eeqs

Third, for a given $N$, $N_f \in I$, and loop order $n=3$ or $n=4$, 
\beq
\alpha'_{IR,n\ell,S_{Q_r}} \quad {\rm is \ a \ decreasing \ function \ of} \ 
r. 
\label{alfir_sqr_r_dependence}
\eeq
%


\section{Comparative Discussion of Scheme Transformations}
\label{comparisons}

\subsection{$S_{sh_r}$ Scheme Transformation} 

It is of interest to compare the $S_{L_r}$ and $S_{Q_r}$
scheme transformations with the the $S_{\rm{sh}_r}$ scheme transformation 
studied in \cite{scc,sch}, 
\beq
S_{\rm{sh}_r}: \quad a=\frac{\sinh(ra')}{r} \ .
\label{aap_sinh}
\eeq
Since $\sinh(ra')/r$ is an even function of $r$, one may take $r \ge 0$ without
loss of generality. Equation (\ref{aap_sinh}) has the inverse
\beq
a' = \frac{1}{r} \, \ln \bigg [ ra + \sqrt{1+ (ra)^2} \ \bigg ] \ . 
\label{sinh_inverse}
\eeq
The corresponding transformation function is
\beq
f(a') = \frac{\sinh(ra')}{ra'} \ , 
\label{fap_sh}
\eeq
with expansion coefficients $k_{s}=0$ for odd $s$ and 
\beq
k_2 = \frac{r^2}{6}, \quad k_4 = \frac{r^4}{120}, \quad k_6 =\frac{r^6}{5040} 
\ ,
\label{ks_sh}
\eeq
etc. for $s \ge 8$. Thus, for small $ra'$, 
\beq
a=a'\Big [ \, 1+ \frac{(ra')^2}{6} +O \Big ( (ra')^4 \Big ) \Big ] \ . 
\label{aaptaylor_sh}
\eeq
The Jacobian is
\beq
J = \frac{da}{da'} = \cosh(ra') \ .
\label{j_sh}
\eeq
This Jacobian always satisfies condition $C_3$. From (\ref{ks_sh}) or
(\ref{j_sh}), it follows that $a' < a$ for nonzero $r$ with this
$S_{\rm{sh}_r}$ scheme transformation.


\subsection{Comparative Discussion of Results with Different Scheme
  Transformations} 

From the studies of a variety of scheme transformations in
\cite{scc}-\cite{sch3} and the present work, a number of general conclusions
follow.  These include the basic properties noted in Eqs. (\ref{alfprime_r0}),
(\ref{alf_lim_Nfb1z}), and the fact that for small $|r|$, the order of the
values of the three-loop and four-loop IR zeros of the beta function are the
same as in the $\overline{\rm MS}$ scheme, (\ref{alfir_loop_dep_msbar}). 

One basic property is that for values of the parameter(s) determining $f(a')$
(here, the parameter $r$ for the $S_{L_r}$, $S_{Q_r}$, and $S_{\rm{sh}_r}$
transformations) such that $f(a')$ does not differ too much from the identity,
the sign of the leading $k_s$ coefficient in the expansion (\ref{fapseries})
determines whether $a'$ is greater or smaller than $a$.  For the $S_{L_r}$
scheme transformation, this leading term for small positive $r$ is negative
(cf. Eq. (\ref{aap_sln})), so $a' > a$, while for the $S_{Q_r}$ and
$S_{\rm{sh}_r}$ scheme transformations, this leading term for small positive
$r$ is positive (cf.  Eqs. (\ref{aap_sqr}) and (\ref{aaptaylor_sh})), so $a' <
a$.  Recall that with the $S_{\rm{sh}_r}$ transformation, the leading term in
the expansion (\ref{fapseries}) is the $k_2(a')^2$ term, while for the
$S_{L_r}$ and $S_{Q_r}$ transformations, the leading term is $k_1(a')$.

In a similar manner, for a general scheme transformation $S_r$, the sign of the
leading correction term in (\ref{fapseries}) also determines whether
$\alpha'_{IR,n\ell,S_r}$ is an increasing or decreasing function of $r$ for
small $|r|$. Thus, the leading correction terms in $S_{L_r}$ scheme
transformation is negative, and $\alpha'_{IR,n\ell,L_r}$ is an increasing
function of $r$, while for the the $S_{Q_r}$ and $S_{\rm{sh}_r}$ scheme
transformations, the leading correction term in (\ref{fapseries}) is positive,
and $\alpha'_{IR,n\ell,S_{Q_r}}$ and $\alpha'_{IR,n\ell,S_{\rm{sh}_r}}$ are
decreasing functions of $r$ and $|r|$, respectively \cite{schmp}.

Concerning the range of $r$ over which a scheme transformation obeys the
conditions $C_1$-$C_4$, we note that for the $S_{\rm{sh}_r}$ transformation
studied in \cite{sch}, this range extended up to at least $|r|=4\pi$, as was
evident from the results displayed in Table III of \cite{sch}.  Here, for the
$S_{L_r}$ and also $S_{Q_r}$ scheme transformations, the respective allowed
ranges of (positive and negative values of) $r$ are somewhat smaller.  This is
easily understood if one examines the Taylor series expansions of the
respective transformation functions $f(a')$.  The values of the coefficients
$k_s$ with even $s$ (the odd-$s$ ones being zero) for the $S_{\rm{sh}_r}$
transformation in Eq. (\ref{ks_sh}) are much smaller than those for the $k_s$
for both the $S_{L_r}$ and $S_{Q_r}$ transformations, listed, respectively, in
Eqs. (\ref{ks_sln}) and (\ref{ks_sqr}).  Therefore, a given value of $r$ leads
to a transformation function $f(a')$ that is considerably closer to the
identity for the $S_{\rm{sh}_r}$ scheme transformation than for the $S_{L_r}$
or $S_{Q_r}$ transformation.  In general, if one constructs and applies a
particular scheme transformation, one can see how large a deviation from the
identity a moderate value of $r$ will produce for the transformation function
$f(a')$ by examining the Taylor series expansion (\ref{fapseries}).


\section{Some Other $S_{[p,q]}$ Scheme Transformations}
\label{other_spq}


\subsection{$S_{[1,1]}$ Scheme Transformation}

In this section we remark on some other $S_{[p,q]}$ scheme transformations with
$q \ne 0$.  We begin with $S_{[1,1]}$, This is defined by the special
case of (\ref{aapspq}) with $[p,q]=[1,1]$, namely 
\beq
S_{[1,1]}: \quad f(a') = \frac{1+u_1a'}{1+v_1a'} \ , 
\label{aap_s11}
\eeq
where $u_1$ and $v_1$ are (real) parameters. The inverse of 
Eq. (\ref{aap_s11}) formally involves two solutions to a quadratic equation,
but only one is physical, because it is the only one for which $a' \to a$ as 
$(u_1,v_1) \to (0,0)$.  This inverse transformation is 
\beq
a' = \frac{-1+v_1a + \sqrt{(1-v_1a)^2 + 4u_1a}}{2u_1} \ . 
\label{apa_s11}
\eeq
The Jacobian is 
\beq
J = \frac{1+2u_1a'+u_1v_1(a')^2}{(1+v_1a')^2} \ . 
\label{j_s11}
\eeq
The transformation function has a Taylor series expansion of the form
(\ref{fapseries}) with 
\beq
k_s = (u_1-v_1)(-v_1)^{s-1} \ . 
\label{ks_s12}
\eeq
%


\subsection{$S_{[1,2]}$ Scheme Transformation}

The $S_{[1,2]}$ scheme transformation is the special case of (\ref{aapspq})
with $[p,q]=[1,2]$, namely
\beq
S_{[1,2]}: \quad f(a') = \frac{1+u_1a'}{1+v_1a'+v_2(a')^2} \ , 
\label{aap_s12}
\eeq
depending on the three (real) parameters $u_1$, $v_1$, and $v_2$. As with
$S_{[1,1]}$, the inverse of (\ref{aap_s12}) formally involves two solutions to
a quadratic equation, but only one is physical because it is the only one for
which $a' \to a$ as $(u_1,v_1,v_2) \to (0,0,0)$.  This inverse transformation
is
\beq
a' = \frac{-1+v_1a + \sqrt{(1-v_1a)^2 + 4a(u_1-v_2a)}}{2(u_1-v_2a)} \ . 
\label{apa_s12}
\eeq
The Jacobian is 
\beq
J = \frac{1+2u_1a'+(u_1v_1-v_2)(a')^2}{(1+v_1a'+v_2(a')^2)^2} \ . 
\label{j_s12}
\eeq
The transformation function has a Taylor series expansion of the form 
(\ref{fapseries}), but with coefficients $k_s$ that are more complicated than
those for $S_{[0,1]}$ or $S_{[1,1]}$.  The first few of these 
coefficients $k_s$ are 
\beq
k_1 = (u_1-v_1) \ ,
\label{k1_s12}
\eeq
\beq
k_2 = -(u_1-v_1)v_1-v_2 \ ,
\label{k2_s12}
\eeq
\beq
k_3 = (u_1-v_1)v_1^2+(2v_1-u_1)v_2 \ ,
\label{k3_s12}
\eeq
\beq
k_4 = -(u_1-v_1)v_1^3+(v_2-3v_1^2+2u_1v_1)v_2 \ ,
\label{k4_s12}
\eeq
and so forth for higher $s$. 

For sufficiently small $|u_1|$ and $|v_1|$, the $S_{[1,1]}$ scheme
transformation obeys the conditions to be applicable at a (perturbatively
calculated) IR zero of the beta function of an asymptotically free gauge
theory.  Similarly, for sufficiently small $|u_1|$, $|v_1|$, and $|v_2|$, 
the $S_{[1,2]}$ scheme also obeys these conditions. Because these scheme
transformations involve two and three parameters, respectively, the analysis of
the allowed ranges of these parameters is more complicated than the
corresponding analyses given in \cite{scc,sch2,sch3} and for the 
one-parameter scheme transformations $S_{L_r}$ and $S_{Q_r}$ here. 

One could also consider $S_{[p,q]}$ scheme transformations with higher (finite)
values of $p$ and/or $q$, but the inverses generically involve equations of
cubic and higher degree, rendering the analytic calculations more
cumbersome. We will thus not pursue these here. 


\section{Infrared-Free Theories}
\label{irfree}

We have focused in this paper on the application of our scheme transformations
$S_{L_r}$ and $S_{Q_r}$ to the study of the scheme dependence of the IR zero of
the beta function in asymptotically free gauge theories.  The question of
scheme dependence also arises in studying the beta function to three loops and
higher in an infrared-free theory, such as (in $d=4$ spacetime dimensions) (i)
a U(1) gauge theory, (ii) a non-Abelian gauge theory with $N_f > N_{f,b1z}$
fermions in a given representation; and (iii) an O($N$) $\lambda
|\vec{\phi}|^4$ scalar field theory.  These IR-free theories have an IRFP at
zero coupling, and one may search for a possible UV zero of the respective beta
function.  Again, it is straightforward to construct acceptable scheme
transformations to apply in the vicinity of the IR fixed point of these
theories at zero gauge or quartic scalar coupling, respectively, but
considerably more difficult to do this when searching for a possible UV zero of
the beta function (UVFP) away from the origin.  Recently this search has been
performed up to five-loop order in \cite{lnf} for theories of type (i) and (ii)
(see also \cite{holdom}) and in \cite{lam} for theories of type (iii), with the
finding of evidence against the existence of such a UVFP in these theories.
Among other methods, these analyses made use of scheme transformations. Since
these findings were quite robust, we have not deemed it necessary to apply the
scheme transformations constructed here to these IR-free theories.  An example
of an IR-free theory that does exhibit such a UV zero (UVFP) was demonstrated
from an exact solution of the O($N$) nonlinear $\sigma$ model in $d=2+\epsilon$
dimensions in the $N \to \infty$ limit \cite{nlsm}.


\section{Conclusions}
\label{conclusions} 

In this paper we have presented two new scheme transformations, $S_{L_r}$ and
$S_{Q_r}$, and have used these to study the scheme-dependence of an infrared
fixed point in an asymptotically free non-Abelian gauge theory, making
comparison with the previous three-loop and four-loop calculations of the
location of this point in the $\overline{\rm MS}$ scheme in \cite{bvh,ps}. Each
of these scheme transformations depends on a parameter $r$, and we have shown
that for a considerable range of values of $r$ in the two respective cases, for
values of the scheme-independent two-loop IR zero of the beta function
$\alpha_{IR,2\ell}$ that are sufficiently small that perturbative calculations
are reasonably reliable, these scheme transformations introduce only relatively
small shifts in the higher-loop values $\alpha_{IR,n\ell,S_{L_r}}$ and
$\alpha_{IR,n\ell,S_{Q_r}}$, as compared with the respective
$\alpha_{IR,n\ell,\overline{\rm MS}}$ for $n=3$ and $n=4$ loops.  This agrees
with and extends the results obtained with the $S_{\rm{sh}_r}$ scheme
transformation in \cite{sch} and also with the results of studies of different
scheme transformations and specific schemes in \cite{scc}-\cite{tr_gen}. Our
results thus provide a further quantitative measure of the size of the
scheme-dependence in the calculation of this fixed point at the three-loop and
four-loop order, both at small and moderate couplings.  We have also 
remarked on a generalized family of multi-parameter scheme transformations,
$S_{[p,q]}$. 


\begin{acknowledgments}

This research was partly supported by the NSF Grant
No. NSF-PHY-13-16617. R. S. thanks T. A. Ryttov for valuable collaborative 
work on the related Refs. \cite{scc,sch} and \cite{bvh,bfs}.

\end{acknowledgments}


\begin{appendix}

\section{Some Relevant Formulas}
\label{appendix}

In this appendix we include some relevant formulas used in the text. 
We first list the beta function coefficients $b_\ell'$
calculated in \cite{scc,sch} that follow from a scheme transformation 
(\ref{aap}), as functions of $b_n$ in the original scheme.  
For our present analysis, we will use the three-loop and
four-loop results \cite{scc,sch} 
\beq
b_3' = b_3 + k_1b_2+(k_1^2-k_2)b_1 \ , 
\label{b3prime}
\eeq
and
\beq
b_4' = b_4 + 2k_1b_3+k_1^2b_2+(-2k_1^3+4k_1k_2-2k_3)b_1 \ .
\label{b4prime}
\eeq
For our analysis of an interesting property of the $S_{Q_r}$ scheme
transformation, we also display $b_5'$: 
\beqs
b_5' & = & b_5+3k_1b_4+(2k_1^2+k_2)b_3+(-k_1^3+3k_1k_2-k_3)b_2 \cr\cr
     & + & (4k_1^4-11k_1^2k_2+6k_1k_3+4k_2^2-3k_4)b_1 \ .
\label{b5prime}
\eeqs

For a vectorial gauge theory with $N_f$ (massless) fermions transforming
according to the representation $R$ of the gauge group $G$, the two
scheme-independent coefficients in the beta function are \cite{b1}
\beq
b_1 = \frac{1}{3}(11 C_A - 4T_fN_f)
\label{b1}
\eeq
and \cite{b2}
\beq
b_2=\frac{1}{3}\left [ 34 C_A^2 - 4(5C_A+3C_f)T_f N_f \right ]
\ .
\label{b2}
\eeq
The calculations of \cite{bvh}, which are used as input for the present
work, used $b_3$ and $b_4$ as calculated in the $\overline{\rm MS}$ scheme
in \cite{b3,b4}. 

\end{appendix}



\newpage

\begin{widetext}


\begin{table}
\caption{\footnotesize{Values of the IR zero, $\alpha'_{IR,3\ell,S_{L_r}}$, of
the three-loop beta function $\beta_{\alpha',3\ell}$ obtained by applying the
$S_{L_r}$ scheme transformation to the three-loop beta function in the
$\overline{\rm MS}$ scheme, for an SU(3) gauge theory with $N_f$ fermions in
the fundamental representation.  For compact notation, we set
$\alpha'_{IR,3\ell,S_{L_r}} \equiv \alpha'_{IR,3\ell,r}$ in the table.  For
each $N_f$, we list these values as a function of $r$ for $r$ from $r=-3$ to
$r=3$ in steps of 1. For $r=0$, $\alpha'_{IR,3\ell,S_{L_r}} =
\alpha_{IR,3\ell,\overline{\rm MS}}$. We also list the (scheme-independent)
two-loop IR zero of the beta function, $\alpha_{IR,2\ell}$.}}
\begin{center}
\begin{tabular}{|c|c|c|c|c|c|c|c|c|} \hline\hline
$N_f$ & $\alpha_{IR,2\ell}$ &
$\alpha'_{IR,3\ell,r=-3}$  &
$\alpha'_{IR,3\ell,r=-2}$  & 
$\alpha'_{IR,3\ell,r=-1}$  &
$\alpha_{IR,3\ell,\overline{MS}}$ & 
$\alpha'_{IR,3\ell,r=1}$  &
$\alpha'_{IR,3\ell,r=2}$  & 
$\alpha'_{IR,3\ell,r=3}$  \\
\hline
10 & 2.21  & 0.749 & 0.754 & 0.759 & 0.764 & 0.769 & 0.774 & 0.778  \\
11 & 1.23  & 0.566 & 0.570 & 0.574 & 0.578 & 0.583 & 0.587 & 0.591  \\
12 & 0.754 & 0.426 & 0.429 & 0.432 & 0.435 & 0.438 & 0.441 & 0.444  \\
13 & 0.468 & 0.311 & 0.313 & 0.315 & 0.317 & 0.319 & 0.321 & 0.323 \\
14 & 0.278 & 0.211 & 0.212 & 0.213 & 0.2145& 0.216 & 0.217 & 0.218 \\
15 & 0.143 & 0.122 & 0.122 & 0.123 & 0.123 & 0.124 & 0.124 & 0.125 \\
16 & 0.0416&0.0396& 0.0396& 0.0397& 0.0397& 0.0398& 0.0398& 0.0399 \\
\hline\hline
\end{tabular}
\end{center}
\label{alfir_3loop_sln}
\end{table}


\begin{table}
\caption{\footnotesize{Values of the IR zero, $\alpha'_{IR,4\ell,S_{L_r}}$, of
the four-loop beta function $\beta_{\alpha',4\ell}$ obtained by applying the
$S_{L_r}$ scheme transformation to the four-loop beta function in the
$\overline{\rm MS}$ scheme, for an SU(3) gauge theory with $N_f$ fermions in
the fundamental representation.  For compact notation, we set
$\alpha'_{IR,4\ell,S_{L_r}} \equiv \alpha'_{IR,4\ell,r}$ in the table.  For
each $N_f$, we list these values as a function of $r$ for $r$ from $r=-3$ to
$r=3$ in steps of 1. For $r=0$, $\alpha'_{IR,4\ell,S_{L_r}} =
\alpha_{IR,4\ell,\overline{\rm MS}}$. We also list 
$\alpha_{IR,2\ell}$ and $\alpha_{IR,3\ell,\overline{\rm MS}}$.}}
\begin{center}
\begin{tabular}{|c|c|c|c|c|c|c|c|c|c|} \hline\hline
$N_f$ & $\alpha_{IR,2\ell}$ & $\alpha_{IR,3\ell,\overline{\rm MS}}$ & 
$\alpha'_{IR,4\ell,r=-3}$  &
$\alpha'_{IR,4\ell,r=-2}$  & 
$\alpha'_{IR,4\ell,r=-1}$  &
$\alpha_{IR,4\ell,\overline{MS}}$ & 
$\alpha'_{IR,4\ell,r=1}$  &
$\alpha'_{IR,4\ell,r=2}$  & 
$\alpha'_{IR,4\ell,r=3}$  \\
\hline
10 & 2.21 &  0.764 & 0.734 & 0.760 & 0.785 & 0.815 & 0.851 & 0.895 & 0.956 \\
11 & 1.23 &  0.578 & 0.576 & 0.591 & 0.607 & 0.626 & 0.648 & 0.673 & 0.705 \\
12 & 0.754 & 0.435 & 0.441 & 0.450 & 0.460 & 0.470 & 0.482 & 0.496 & 0.511 \\
13 & 0.468 & 0.317 & 0.322 & 0.327 & 0.332 & 0.337 & 0.343 & 0.349 & 0.356 \\
14 & 0.278 & 0.2145& 0.217 & 0.219 & 0.221 & 0.224 & 0.226 & 0.228 & 0.231 \\
15 & 0.143 & 0.123 & 0.124 & 0.124 & 0.125 & 0.126 & 0.126 & 0.127 & 0.128 \\
16 & 0.0416&0.0397& 0.0396& 0.0397& 0.0398& 0.0398& 0.0399& 0.0400& 0.0400 \\
\hline\hline
\end{tabular}
\end{center}
\label{alfir_4loop_sln}
\end{table}


\begin{table}
\caption{\footnotesize{Values of the IR zero, $\alpha'_{IR,3\ell,S_{Q_r}}$, of
the three-loop beta function $\beta_{\alpha',3\ell}$ obtained by applying the
$S_{Q_r}$ scheme transformation to the three-loop beta function in the
$\overline{\rm MS}$ scheme, for an SU(3) gauge theory with $N_f$ fermions in
the fundamental representation.  For compact notation, we set
$\alpha'_{IR,3\ell,S_{Q_r}} \equiv \alpha'_{IR,3\ell,r}$ in the table.  For
each $N_f$, we list these values as a function of $r$ for $r$ from $r=-3$ to
$r=3$ in steps of 1. For $r=0$, $\alpha'_{IR,3\ell,S_{Q_r}} =
\alpha_{IR,3\ell,\overline{\rm MS}}$. We also list the 
(scheme-independent) two-loop value of the IR zero, $\alpha_{IR,2\ell}$.}}
\begin{center}
\begin{tabular}{|c|c|c|c|c|c|c|c|c|} \hline\hline
$N_f$ & $\alpha_{IR,2\ell}$ &
$\alpha'_{IR,3\ell,r=-3}$  &
$\alpha'_{IR,3\ell,r=-2}$  & 
$\alpha'_{IR,3\ell,r=-1}$  &
$\alpha_{IR,3\ell,\overline{MS}}$ & 
$\alpha'_{IR,3\ell,r=1}$  &
$\alpha'_{IR,3\ell,r=2}$  & 
$\alpha'_{IR,3\ell,r=3}$  \\
\hline
10 & 2.21 &  0.795 & 0.785 & 0.774 & 0.764 & 0.755 & 0.746 & 0.737  \\
11 & 1.23 &  0.605 & 0.596 & 0.587 & 0.5785& 0.571 & 0.563 & 0.556  \\
12 & 0.754 & 0.455 & 0.448 & 0.441 & 0.435 & 0.429 & 0.423 & 0.418  \\
13 & 0.468 & 0.330 & 0.325 & 0.321 & 0.317 & 0.313 & 0.309 & 0.305  \\
14 & 0.278 & 0.222 & 0.219 & 0.217 & 0.215 & 0.212 & 0.210 & 0.208  \\
15 & 0.143 & 0.126 & 0.125 & 0.124 & 0.123 & 0.122 & 0.122 & 0.121  \\
16 & 0.0416&0.0401& 0.0400& 0.0398& 0.0397& 0.0396& 0.0395& 0.0394  \\
\hline\hline
\end{tabular}
\end{center}
\label{alfir_3loop_sqr}
\end{table}


\begin{table}
\caption{\footnotesize{Values of the IR zero, $\alpha'_{IR,4\ell,S_{Q_r}}$, of
the four-loop beta function $\beta_{\alpha',4\ell}$ obtained by applying the
$S_{Q_r}$ scheme transformation to the four-loop beta function in the
$\overline{\rm MS}$ scheme, for an SU(3) gauge theory with $N_f$ fermions in
the fundamental representation.  For compact notation, we set
$\alpha'_{IR,4\ell,S_{Q_r}} \equiv \alpha'_{IR,4\ell,r}$ in the table.  For
each $N_f$, we list these values as a function of $r$ for $r$ from $r=-3$ to
$r=3$ in steps of 1. For $r=0$, $\alpha'_{IR,4\ell,S_{Q_r}} =
\alpha_{IR,4\ell,\overline{\rm MS}}$. We also list 
$\alpha_{IR,2\ell}$ and 
$\alpha_{IR,3\ell,\overline{\rm MS}}$. The dash notation ($-$) means that the 
transformation yields an unphysical (here, complex) value for
$\alpha'_{IR,4\ell,S_{Q_r}}$.}}
\begin{center}
\begin{tabular}{|c|c|c|c|c|c|c|c|c|c|} \hline\hline
$N_f$ & $\alpha_{IR,2\ell}$ & $\alpha_{IR,3\ell,\overline{\rm MS}}$ & 
$\alpha'_{IR,4\ell,r=-3}$  &
$\alpha'_{IR,4\ell,r=-2}$  & 
$\alpha'_{IR,4\ell,r=-1}$  &
$\alpha_{IR,4\ell,\overline{MS}}$ & 
$\alpha'_{IR,4\ell,r=1}$  &
$\alpha'_{IR,4\ell,r=2}$  & 
$\alpha'_{IR,4\ell,r=3}$  \\
\hline
10 & 2.21 &  0.764 & $-$   & 1.062 & 0.896 & 0.815 & 0.760 & 0.719 & 0.685 \\
11 & 1.23 &  0.578 & $-$   & 0.750 & 0.674 & 0.626 & 0.591 & 0.563 & 0.540 \\
12 & 0.754 & 0.435 & 0.581 & 0.530 & 0.496 & 0.470 & 0.450 & 0.433 & 0.418 \\
13 & 0.468 & 0.317 & 0.380 & 0.363 & 0.349 & 0.337 & 0.327 & 0.318 & 0.309 \\
14 & 0.278 & 0.2145& 0.239 & 0.233 & 0.228 & 0.224 & 0.219 & 0.215 & 0.211 \\
15 & 0.143 & 0.123 & 0.130 & 0.128 & 0.127 & 0.126 & 0.124 & 0.123 & 0.122 \\
16 & 0.0416&0.0397& 0.0402& 0.0401& 0.0400& 0.0398& 0.0397& 0.0396& 0.0395 \\
\hline\hline
\end{tabular}
\end{center}
\label{alfir_4loop_sqr}
\end{table}

\end{widetext} 



\begin{thebibliography}{99}
\bibitem{casimir}
%
The Casimir invariants $C_R$ and $T_R$ are defined as $\sum_a \sum_j {\cal
D}_R(T_a)_{ij} {\cal D}_R(T_a)_{jk} = C_R \delta_{ik}$ and $\sum_{i,j} {\cal
D}_R(T_a)_{ij} {\cal D}_R(T_b)_{ji} = T_R \delta_{ab}$, where $R$ is the
representation and $T_a$ are the generators of $G$, so that for SU($N_c$),
$C_A=N_c$ for the adjoint ($A$) and $T_{fund}=1/2$ for the
fundamental representation, etc. $C_f$ denotes $C_R$ for the fermion 
representation.
\bibitem{nfreal}
%
Here and elsewhere, when an expression is given for $N_f$ 
that formally evaluates to a non-integral real value, it is understood
implicitly that one infers an appropriate integral value from it, either the
greatest integer smaller than the real value, or the least integer greater than
the real value, or the closest integer, depending on the context.
\bibitem{schmp}
%
In Eqs. (7.13) and (7.15) of Ref. \cite{sch}, in the subscripts $S_{{\rm
th}_r}$ should read $S_{{\rm sh}_r}$, as in Eq. (7.14) of Ref. \cite{sch}.
\end{thebibliography}
\end{document}